# Near-single-domain superconducting aluminum films on GaAs(111)A with exceptional crystalline quality for scalable quantum circuits

Hsien-Wen Wan[1,#], Yi-Ting Cheng[1,#], Chao-Kai Cheng[1,2], Jui-Min Chia[3], Chien-Ting Wu[4], Sheng-Shiuan Yeh[5], Chia-Hung Hsu[2], Jueinai Kwo[3,*], Minghwei Hong[1,*]

[1] Grad. Institute of Applied Physics and Dept. of Physics, National Taiwan Univ., Taipei, Taiwan

[2] National Synchrotron Radiation Research Center, Hsinchu, Taiwan

[3] Department of Physics, National Tsing Hua University, Hsinchu, Taiwan

[4] Taiwan Semiconductor Research Institute, Hsinchu, Taiwan

[5] Intl. College of Semiconductor Technology, Natl. Yang Ming Chiao Tung Univ., Hsinchu, Taiwan

[#] These authors contributed equally to this work.

* Authors to whom the correspondence should be addressed. Electronic addresses:
   raynien@phys.nthu.edu.tw (J. Kwo), and mhong@phys.ntu.edu.tw (M. Hong).

## Abstract

We have reproducibly grown near-single-domain superconducting aluminum (Al) films on GaAs(111)A wafers using molecular beam epitaxy. Synchrotron X-ray diffraction revealed twin-domain ratios of 0.00005 and 0.0003 for 19.4-nm- and 9.6-nm-thick films, respectively—the lowest reported for Al on any substrate and long considered unattainable for practical device platforms. Azimuthal scans across off-normal Al$\{11\bar{1}\}$ reflections exhibit narrow full width at half maximum (FWHM) values down to 0.55°, unmatched by *epi*-Al grown by any other method. Normal scans showed a well-defined (111) orientation with pronounced Pendellösung fringes, and θ-rocking-curve FWHM values down to 0.018°; the former indicates abrupt film–substrate and oxide–film interfaces. Electron backscatter diffraction mapping confirms macroscopic in-plane uniformity and the absence of Σ3 twin domains. Atomic force microscopy and scanning transmission electron microscopy confirmed atomically smooth surfaces and abrupt heterointerfaces. The films exhibit critical temperatures approaching bulk values, establishing a materials platform for scalable, high-coherence superconducting qubits.

**Introduction**

Quantum computing stands at the forefront of transformative technologies, offering the potential to solve problems far beyond the reach of classical computation. Among the various physical implementations, superconducting qubits, built from Josephson junctions (JJs) and microwave resonators, have emerged as the most advanced and scalable platform. The JJs of these qubits, typically based on native aluminum oxide ($AlO_x$)/aluminum (Al) heterostructures, have achieved coherence times exceeding one millisecond[1-3]. Most recently, Google demonstrated quantum error correction below the surface code threshold, showing that logical qubit performance improves with increasing code distance, a landmark toward fault-tolerant quantum computation[4].

The performance of superconducting qubits depends critically on the quality of their constituent materials. Conventional $AlO_x$/Al heterostructures are produced by thin-film deposition, in which Al is deposited by e-beam evaporation onto sapphire or silicon (Si) substrates and subsequently oxidized to form the tunnel barrier. These Al films are typically polycrystalline or epitaxial with nearly equal twin-domain populations and granular morphologies. The associated grain and twin boundaries act as diffusion channels for oxygen and contaminants during device fabrication and air exposure[5-9]. Such structural and chemical imperfections introduce defects within the films and on their surfaces that limit qubit coherence and device stability[10-13]. Achieving high-coherence superconducting qubits, therefore, requires epitaxial Al films with minimal grain and twin boundaries, ensuring structural robustness, chemical uniformity, and suppression of defect-induced dielectric losses.

Scaling quantum processors to the millions of physical qubits required for error-corrected computation[4] imposes stringent demands on uniformity and reproducibility of defect-free material. Lessons from semiconductor technology underscore the decisive role of single-crystal Si substrates, whose atomically smooth, defect-free lattices underpinned the scalability of complementary metal-oxide-semiconductor (CMOS) electronics. In contrast, superconducting circuits rely on thin-film superconductors, where realizing single-crystal-like perfection via heteroepitaxy remains challenging. The difficulty arises from lattice mismatch, interfacial chemical bonding disparities, and defect formation at interfaces between dissimilar materials.

Historically, breakthroughs in heteroepitaxial materials engineering have catalyzed major technological advances. The development of GaN-on-sapphire enabled solid-state lighting[14]; epitaxial Gd/Y magnetic superlattices on Nb/sapphire enabled discoveries of long-range magnetic and antiferromagnetic coupling[15,16], leading to the observation of giant magnetoresistance (GMR)[17,18]; and epitaxial growth of rare-earth oxides on GaAs[19,20], InGaAs[21-23], and GaN[24], unpinned Fermi-levels in compound semiconductors. Each case illustrates how the perfection of interfacial crystallinity and bonding drives the emergence of new device paradigms.

Epitaxial Al thin films have been studied on a variety of substrates, including GaAs(100)[25-28], sapphire[5,6,8,28,29], Si(100)[30,31], and Si(111)[28,30-35]. For Al films on GaAs(100), film orientation depends on surface reconstruction, growth rate, and substrate temperature: (110) on As-rich surfaces and (100) on Ga-rich surfaces[25,26], with occasional (111) orientation on GaAs(100)-(4×6) Ga-rich surfaces[27,28]. Epitaxial Al has also been grown on c-plane sapphire (0001) using e-beam evaporation[13] or molecular beam epitaxy (MBE)[5,6,8,28,29], despite the large lattice mismatch. However, this typically results in equal twin populations and granular morphologies.

On Si(111), the lattice mismatch between Al (4.05 Å) and Si (5.43 Å) can be partially compensated by a 4:3 superlattice arrangement. MBE growth of Al on Si(111)-$\sqrt{3}\times\sqrt{3}$ achieved a twin ratio of 0.001 for 100 nm films[34]. Note that preparing the Si(111)-$\sqrt{3}\times\sqrt{3}$ surface requires complex multi-step procedures, including the deposition of 1/3 monolayer Al on Si(111)-7×7 and subsequent annealing steps at 775°C.[34]

In this work, we used MBE to grow near-single-domain superconducting Al films on GaAs(111)A substrates, which exhibit a (2×2) surface reconstruction consistent with the surface Ga vacancy model[36-38]. GaAs(111)A was chosen based on its proven ability to support epitaxial metal and oxide growth with exceptional structural integrity - e.g., $Fe_3Si$(111) films exhibiting ultra-low surface (0.13 nm) and interfacial (0.19 nm) roughness[39], single-crystal yttrium aluminum perovskite (YAP)[40,41], and $C_{60}$ films[42].

Our Al films exhibit record-low twin-domain ratios of 0.00005 and 0.0003 for 19.4 nm and 9.6 nm thick films, respectively - achieving an unprecedented level of crystalline perfection in epitaxial Al. Synchrotron X-ray diffraction (S-XRD) reveals exceptionally narrow azimuthal and θ-rocking-curve FWHM, along with pronounced Pendellösung fringes, confirming high crystallinity and interface abruptness. Large-area electron backscatter diffraction (EBSD) mapping over hundreds of micrometers directly reveals a single-domain structure with uniform in-plane orientation and an absence of Σ3{111} twin-related variants across the film. Atomic force microscopy (AFM) and cross-sectional scanning transmission electron microscopy (STEM) demonstrate atomically smooth surfaces and sharp interfaces.

In addition to structural quality, these Al films display superconducting properties approaching those of bulk aluminum, including high residual resistivity ratios (RRR) and critical temperatures close to bulk values, surpassing all previously reported films of comparable thickness.

These results establish a new materials paradigm for superconducting aluminum, achieving near-single-domain crystallinity, atomically smooth interfaces, and wafer-scale compatibility comparable to single-crystal semiconductor substrates. This advance provides a materials foundation for the next generation of high-coherence, scalable superconducting qubits, addressing one of the central challenges in realizing practical quantum computing.

## Results and Discussion

**Twin-domain ratios and in-plane domain structure**.

Azimuthal $\phi$-scans across off-normal Al$\{11\bar{1}\}$ reflections (Fig. 1a) display three sharp peaks separated by 120°, reflecting the threefold symmetry of Al$\{11\bar{1}\}$ about the [111] axis. The azimuthal positions coincide precisely with those of the GaAs substrate $\{11\bar{1}\}$ reflections (i.e., at 60°, 180°, and 300°), identifying them as type-A domains with the epitaxial orientation relationship:

$$\text{Al}(111)[2\bar{1}\bar{1}] \parallel \text{GaAs}(111)[2\bar{1}\bar{1}]$$

A second, much weaker set of peaks, rotated by 180° relative to the type-A domains, corresponds to type-B twin domains, characterized by the relationship:

$$\text{Al}(111)[\bar{2}11] \parallel \text{GaAs}(111)[2\bar{1}\bar{1}]$$

Due to their extremely low intensity, the type-B peaks are invisible on the current scale used in Fig. 1a but become discernible after amplifying the peak intensities in the $\phi$ of 240° by $10^4$ (19.4 nm film) and $10^3$ (9.6 nm film) (Fig. 1b). Even after amplification, the type-B signal of the thicker film remains near the background noise level. Fine $\phi$-scans confirm the presence of a faint but measurable peak, verifying a minute population of type-B twins.

The twin-domain ratios, defined as ($A_{\text{type-B}}/A_{\text{type-A}}$), where $A_{\text{type-A}}$ and $A_{\text{type-B}}$ are the peak areas (or integrated intensities) of the Al$\{111\}$ reflections for type-A and type-B domains, respectively, are 0.00005 for the 19.4 nm film and 0.0003 for the 9.6 nm film - the lowest reported to date for Al films on any substrate. These values confirm that the Al/GaAs(111)A heterostructure yields near single-domain Al films, approaching single-crystal characteristics over macroscopic scales through a predominant type-A cube-on-cube epitaxial relationship. These exceptionally low twin ratios have been consistently reproduced in our other epitaxial Al films grown on GaAs(111)A. The detection of the extremely weak domain B peaks, caused by a negligibly small population of domain B, was made possible by the high brilliance of synchrotron radiation.

**Comparison of twin domains and in-plane orientation among films grown on various substrates.**

Table 1 compares the crystalline parameters of Al films grown on GaAs(111)A, Si(111), GaAs(100), and sapphire(0001). E-beam evaporated or MBE Al films on sapphire (0001) (c-plane) exhibit epitaxy and twin domains with equal populations (a twin ratio of 1), as studied using XRD azimuthal scans across off-normal reflections; the epitaxial orientation relationships are $Al(111)[2\bar{1}\bar{1}]$ ∥ $sapphire(0001)[2\bar{1}\bar{1}0]$ and $Al(111)[\bar{2}11]$ ∥ $sapphire(0001)[2\bar{1}\bar{1}0]$, where the twin domains are rotated 180° relative to each other.

Epitaxial Al films on Si(111)-$\sqrt{3}\times\sqrt{3}$) surfaces achieved twin ratios as low as 0.001 for 100 nm thick films, and 0.023 on Si(111)-7×7 surfaces for 20nm. Al films on GaAs(100) are reported to exhibit a twin ratio of 0.742.

Our growth of an Al/GaAs(111)A heterostructure effectively suppresses twin formation, setting a new benchmark for achieving high-quality, single-domain Al thin films. Twin domains facilitate the diffusion of oxygen, hydrocarbon species (C-H containing species), and other species along the boundaries, leading to contamination and native oxide formation within the films and on their surfaces. These structural imperfections are potential decoherence sources that can significantly impact the performance and the long-term stability of quantum devices. The substantial reduction in twin domain boundaries results in lower defect densities, critical for achieving high crystalline perfection for minimizing variability across the wafer and enhancing the performance of superconducting quantum devices.

Additionally, the FWHM values of azimuthal $\phi$-scans across Al$\{11\bar{1}\}$ reflections for domain A were measured to be 0.55° for the 19.4 nm film and 0.76° for the 9.6 nm film. These small FWHM values indicate small lattice twist angles, highly-ordered in-plane crystalline alignment, and reduced structural disorder. Notably, these values are significantly lower than the FWHM values of 2.0°-2.4° reported in previous studies of Al films grown on Si(111), GaAs(100), and sapphire(0001), further demonstrating the exceptional structural quality of the Al films grown in this work. Please see the results as listed in Table 1 for comparison.

**Out-of-plane crystal structure**

S-XRD radial scans along the surface normal (Fig. 1c) exhibit a prominent Al(111) reflection at a scattering vector q = 2.67 Å$^{-1}$, corresponding to the reciprocal lattice vector of Al(111). The abscissa represents the scattering vector, q, calculated as $4\pi \times \sin(2\theta/2)/\lambda$, where $2\theta$ is the scattering angle, and $\lambda$ is the X-ray wavelength. In addition to the GaAs(111) Bragg peak, the presence of only the Al(111) peak confirms the (111)-oriented growth of the Al films. The lattice constants extracted from the Al(111) peak positions (assuming a cubic lattice) match the bulk Al value of 0.405 nm, indicating that the films are unstrained, which is consistent with *in-situ* RHEED observations of Al growth from the initial stage to thicker films. (See Fig. S1 in supplementary information.)

The pronounced Pendellösung fringes surrounding the Al(111) reflections provide compelling evidence of the films' high crystallinity and atomically abrupt interfaces between Al, the GaAs substrate, and the *in-situ* deposited Al$_2$O$_3$ passivation layer. The film thicknesses derived from the fringe periodicities are 9.6 nm and 19.4 nm, which agree closely with the coherence lengths obtained from the Scherrer equation (9.98 nm and 19.6 nm, respectively). The *in-situ* Al$_2$O$_3$ deposition effectively preserves pristine Al surfaces, preventing native oxide formation and contamination.

The full width at half maximum (FWHM) of the θ-rocking curves of the Al(111) reflection provides a quantitative measure of crystalline perfection. The FWHM values range from 0.018° to 0.024° (as shown in Fig. 1d), significantly narrower than those reported for Al films grown on Si(111) or GaAs(100) substrates (Table 1)[28,34]. These small values indicate minimal lattice tilt and highly-ordered crystalline alignment along the growth direction.

Table 1 compares the FWHM values of the θ-rocking curves in the X-ray radial scans along the surface normal of Al films deposited on various substrates, including Si(111)[28,34], GaAs(100)[28], sapphire(0001)[29], and GaAs(111)A. A distinct difference in film quality is observed across different substrates. The θ-rocking FWHM values provide insights into the degree of crystal alignment in the surface-normal direction.

Among the Al films listed in Table 1, those grown on GaAs(100) exhibit the largest FWHM values (~0.31°), reflecting significant defect densities and lower crystalline quality as a result of the different in-plane symmetry. Note that the Al films included in Table 1 were selected because of their high crystalline quality compared to most published results. For Al films on Si(111), the reported rocking curve FWHM values of 0.042° and 0.066° are lower than those observed for GaAs(100), indicating fewer defects.

The Al films grown on GaAs(111)A in this study exhibit significantly lower FWHM values of 0.018° and 0.024°, demonstrating substantially improved crystalline quality. Note that Al films grown on sapphire(0001) also exhibit very low FWHM values (0.015°), indicating high crystalline alignment in the normal direction. However, as discussed in the previous section of S-XRD azimuthal scans, the unavoidable formation of equally populated twin domains in Al films deposited on sapphire(0001) introduces twin boundaries, which are crystal defects.

Furthermore, the X-ray beam used in this study illuminates a relatively large area of the film, on the order of mm$^2$, demonstrating that the high crystallinity observed is not confined to localized regions but instead represents the uniform crystalline quality of the Al thin films across a wide area.

**Electron backscatter diffraction (EBSD)**

Large-area EBSD mapping over an 800 μm × 800 μm area was performed on Al films to examine the grain distribution and twin density. In Fig. 1e, we showed the inverse pole figure (IPF) maps obtained from the 19.4-nm Al film. The Al film exhibits a (111) out-of-plane texture (IPF-Z), which is consistent with the XRD normal-scan results. The natural cleave of the GaAs(111)A substrate was intentionally aligned parallel to the x-axis of the EBSD sample stage, corresponding to the ⟨112⟩ in-plane direction. The IPF-X and IPF-Y maps of the Al film further confirmed well-aligned in-plane orientations along the ⟨112⟩ and ⟨110⟩ directions, respectively, in agreement with the in-plane alignment between the Al film and the GaAs(111)A substrate as revealed by the XRD off-normal $\phi$-scan results.

The pole figures of {111}, {110}, and {112} planes for Al films with three different textures are compared in Fig. 1f. The projection plane is the sample surface plane. The {111} pole figures for Al (111) film confirm the formation of a strong {111} texture. The {111} pole figure shows only the three expected ⟨111⟩ poles, with no additional peaks rotated by 180°, thereby confirming the absence of Σ3{111} twin domains over a large area. The single-domain property of Al(111) is also revealed in all other pole figures, including {110} and {112}. These results are consistent with the XRD off-normal $\phi$-scans of the Al(111) reflections, which exhibit exceptionally sharp single peaks with no twin-related features.

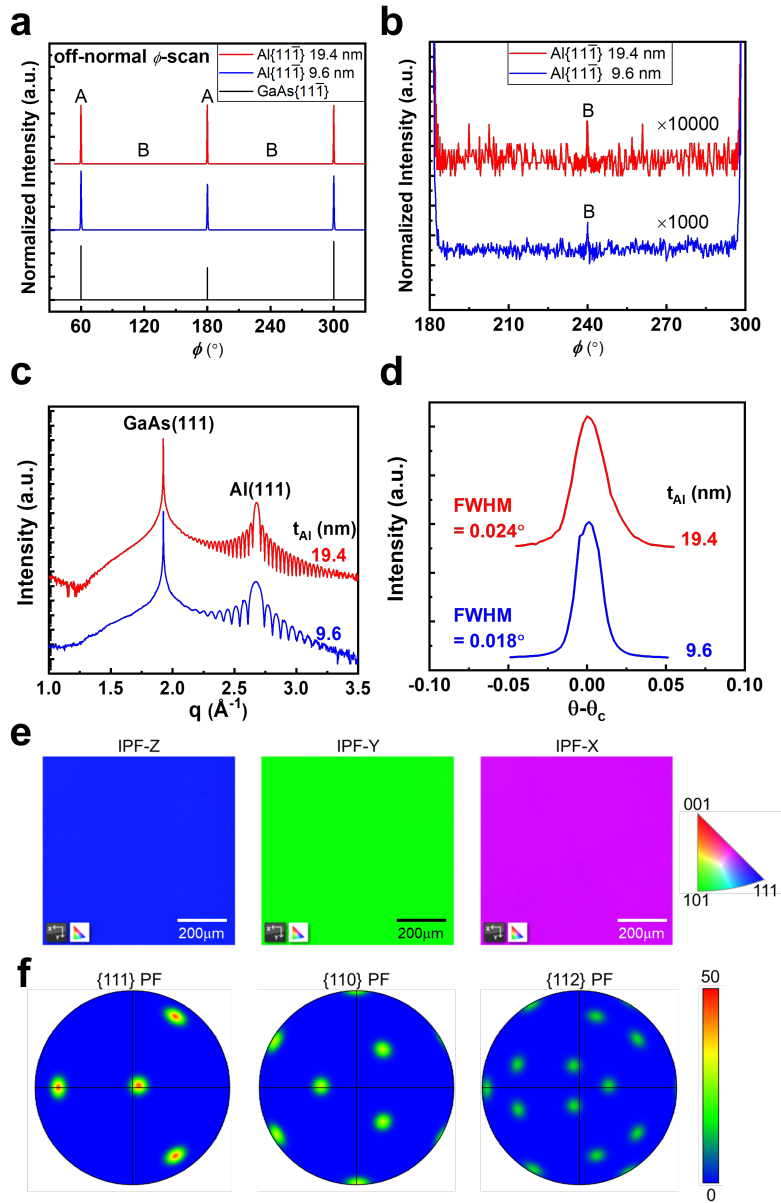

**Fig. 1 | Structural Characterization of Al films grown on GaAs(111)A. a**, Synchrotron X-ray diffraction (S-XRD) $\phi$-scans across off-normal reflections showing single-domain characteristics and **b**, Intensity amplification to enhance the visibility of type-B domain peaks in the $\phi$ range of 180° - 300°, with a 10,000-fold amplification for the 19.4 nm Al peak and a 1,000-fold amplification for the 9.6 nm Al peak. The twinned peaks are labeled as A and B. The twin ratio is calculated as the area (i.e., integrated intensity) of the type-B domain peaks divided by that of the type-A domain peaks. **c**, S-XRD radial scans along the surface normal of 9.6 nm, 19.4 nm-thick Al films. **d**, $\theta$-rocking curves of the Al(111) reflection of both films. **e**, The inverse pole figure (IPF) maps in the sample axis of Z, Y, and X analyzed from the electron backscatter diffraction (EBSD) results. **f**, {111}, {110}, and {112} pole figures of Al film analyzed from the EBSD results. The projected plane is the sample surface.

**Table 1** | Summary of twin-domain ratio, full width at half-maximum (FWHM) of the $\phi$-scans across Al$\{11\bar{1}\}$ reflections, and FWHM in the $\theta$-rocking curves across the Al(111) peak of epitaxial aluminum films on various substrates (NA (not available), #extracted from off-normal $\phi$-scan across Al$\{002\}$.)

| Substrate | Thickness | Twin ratio ($\frac{A_{type-B}}{A_{type-A}}$) | FWHM (off-normal $\phi$-scan across Al$\{11\bar{1}\}$ of Al(111) film) | FWHM ($\theta$-rocking curves of Al(111) peak/sub. normal reflection) | Reference |
|---|---|---|---|---|---|
| GaAs(111)A | 19.4 nm | **0.00005** | **0.55°** | 0.024°/0.006° | This work |
| GaAs(111)A | 9.6 nm | **0.0003** | **0.76°** | 0.018°/0.006° | This work |
| Si(111)-$\sqrt{3}\times\sqrt{3}$ | 100 nm | 0.001 | NA | 0.042°/NA | 34 |
| Si(111)-7x7 | 20 nm | 0.023 | #2.0° | 0.066°/NA | 28 |
| GaAs(100) | 40 nm | 0.742 | #2.4° | 0.31°/NA | 28 |
| Sapphire(0001) | 20 nm | 1 | 2.4° | 0.015°/0.014° | 29 |

**High-angle annular dark-field scanning transmission electron microscopy (HAADF-STEM) Analysis**

Fig. 2 presents a cross-sectional high-angle annular dark-field scanning transmission electron microscopy (HAADF-STEM) image of the Al/GaAs(111)A interface viewed along the [1$\bar{1}$0] direction. The HAADF-STEM image reveals a well-ordered and atomically sharp Al/GaAs(111)A hetero-interface, which is essential for achieving high-quality epitaxial growth. The interface accommodates the lattice mismatch through a periodic atomic arrangement, where every six Ga atoms correspond to eight Al atoms. A detailed comparison of atomic distances along the [11$\bar{2}$] direction indicates that the effective lattice mismatch is minimized to approximately 0.28% when considering a seven Al–Al to five Ga–Ga spacing ratio, ensuring minimal residual strain beyond the interface. The lattice schematics in Fig. 2 illustrate the stacking sequence of Al on GaAs, with blue, orange, and green spheres representing Al, Ga, and As atoms, respectively. The observed atomic configuration in the HAADF-STEM image closely aligns with these schematics, confirming that most of the strain is accommodated at the interface. This is supported by AFM measurements revealing an atomically smooth surface morphology for the Al films grown on GaAs(111)A. (See Fig. S2 in supplementary information.)

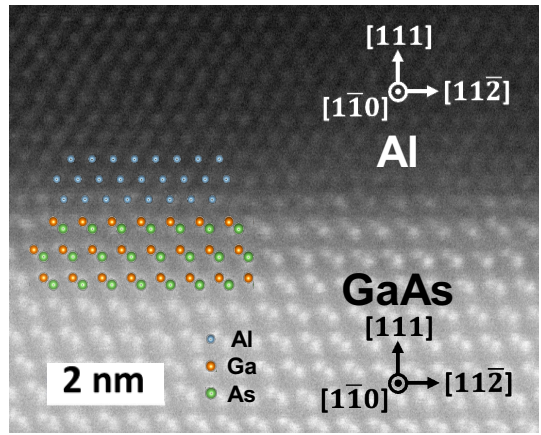

**Fig. 2| High-angle annular dark-field scanning transmission electron microscopy (HAADF-STEM) image of MBE-grown Al(111) film on GaAs(111)A** Cross-sectional STEM image shows an atomically sharp interface. Atomic models overlaid on the HAADF-STEM image indicate the positions of aluminum (blue), gallium (orange), and arsenic (green).

## Resistivity measurements from 300 K to 2 K

Resistivity measurements of the Al films were conducted using a standard four-terminal Hall-bar geometry, with resistance measured in the linear regime of the DC I-V curve. Fig. 3a shows the resistivity of Al films with different thicknesses at 300 K and 2 K. All measured films exhibited higher resistivity at 300 K compared to bulk Al (~2.73 x $10^{-8}$ Ω-m)[43]. An inverse relationship between film thickness and resistivity was observed, likely due to increased surface scattering in thinner films. Fig. 3b displays resistance versus temperature curves for Al films of different thicknesses, showing that the inverse thickness-resistivity relationship persists across the entire temperature range (300 K to 2K). The 9.6 nm-thick film exhibited higher resistivity than the 19.4 nm-thick film, despite both having excellent crystallinity and near-single-domain structures.

The observed behavior is primarily attributed to dimensional effects in thin films. As film thickness decreases, the mean free path of conduction electrons shortens, leading to enhanced electron scattering at surfaces and interfaces[44-46]. In ultra-thin films, the increased surface-to-volume ratio further amplifies electron-surface interactions, significantly reducing charge carrier mobility and modifying the film's electrical properties. These experimental findings align closely with the Fuchs-Sondheimer model, which describes surface scattering effects on electrical resistivity in thin films[47,48].

The residual resistance ratio (RRR) is defined as the ratio of a metal's resistivity at room temperature to its resistivity at low temperatures, such as 2 K. A high RRR signifies high purity and low defect density, as impurities and structural defects (like vacancies, dislocations, and twin- and grain-boundaries) increase low-temperature resistivity, reducing the RRR. Fig. 3c exhibits RRR versus thickness of the 9.6 nm and the 19.4 nm thick Al films grown on GaAs(111)A, exhibiting higher RRR values in the listed thickness range compared with Al films grown on other substrates[29,49,50].

The resistivity measurements provide critical insights into the electronic transport properties of these near-single-domain epitaxial Al films. The observed thickness-dependent behavior and high RRR values highlight the high quality of the films grown on GaAs(111)A, demonstrating their potential for superconducting device applications.

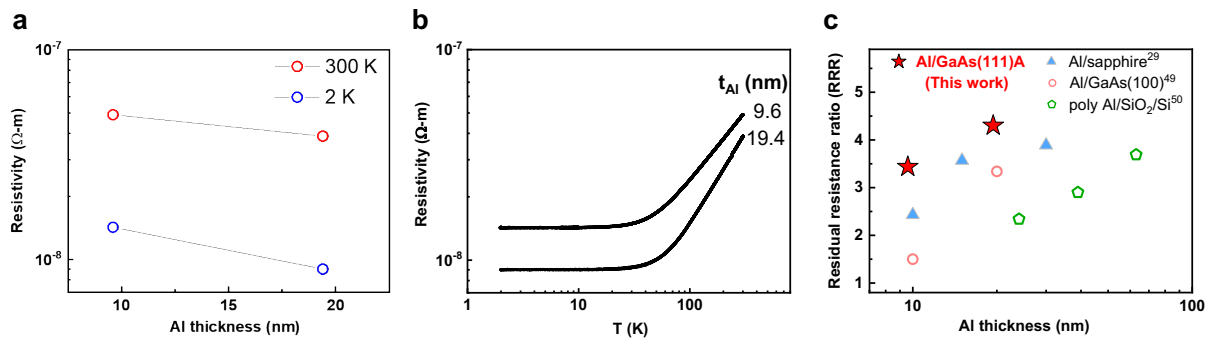

**Fig. 3 | Transport properties of the Al films on GaAs(111)A. a**, Normal-state resistivity measured at 300 K and 2 K of the 9.6 nm and 19.4 nm thick Al films. **b**, Resistivity-temperature (R-T) curves for the respective Al films measured from 300 K to 2 K. **c,** Residual resistance ratio (RRR) versus thickness of Al films grown on GaAs(111)A, exhibiting higher RRR values in the listed thickness range compared with Al films grown on other substrates.

## Superconducting critical temperature ($T_c$)

We measured the superconducting transition curves of Al films grown on GaAs(111)A substrates, revealing several key insights. Sharp transitions from the normal state to the superconducting state were observed for both Al films, as shown in Fig. 4a. The critical temperature ($T_c$) was 1.14 K for the 19.4 nm film and 1.27 K for the 9.6 nm film, both close to the $T_c$ of bulk single-crystal Al (~1.21 K). The $T_c$ behaviors in these two thin films reflect the films' exceptional material quality, characterized by very few defects within the films, such as twin boundaries, and also by minimal effects caused by inevitable boundary interferences from the bottom interface with the GaAs substrate and the top interface with the *in-situ* deposited $Al_2O_3$.

Al films grown on sapphire(0001)[28,29,49] and GaAs(100)[28,49] exhibit twin-domain structures, leading to inferior crystal quality compared to our near-single-domain Al films on GaAs(111)A, as summarized in Table 1. In Fig. 4b, we compare our $T_c$ versus Al thickness data with previous studies, which reported higher $T_c$ values. A particularly relevant comparison comes from our group's recent study on Al films grown on sapphire(0001)[29] using the same growth system and fabrication procedure. While the sapphire(0001)-grown films feature equal populations of twin domains, our GaAs(111)A-grown films exhibit a near-single-domain structure. Despite structural differences, other material properties remain comparable. Notably, our near-single-domain Al films achieve $T_c$ values closer to the bulk, highlighting their high crystallinity and reduced defect density.

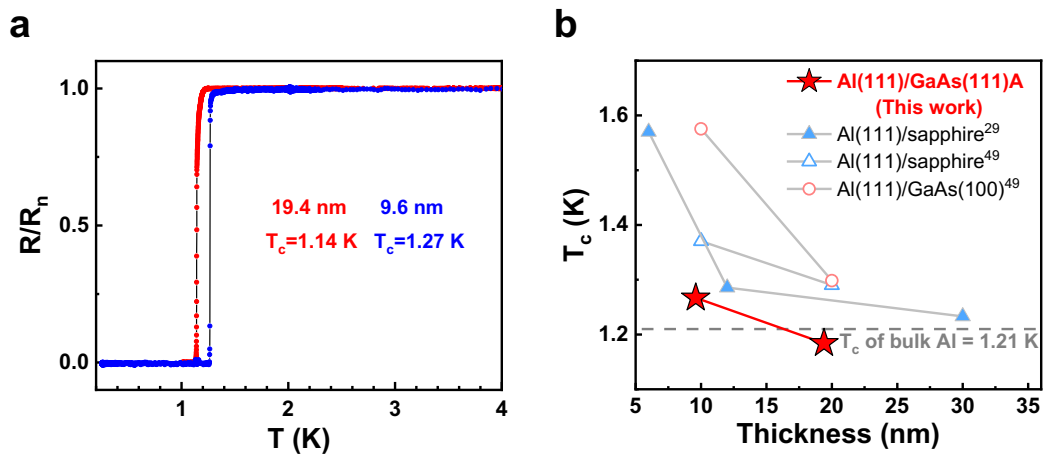

**Fig. 4 | Superconducting transitions and critical temperatures. a**, $R/R_n$ versus temperature curves showing the superconducting transition of Al films grown on GaAs(111)A. **b**, Comparison of $T_c$ values in this work with those from other research efforts, with our Al/GaAs(111)A samples exhibiting the lowest $T_c$ values.

## Conclusion

This work establishes that superconducting aluminum films - the foundational material of state-of-the-art quantum processors - can be elevated from a polycrystalline or epitaxial twined thin film to a near-single-crystal material platform, thus eliminating grain- or twin-boundary-dominated disorder at the wafer scale. The electrical transport properties achieved in this work enable our aluminum films to have high potential for applications in large-scale qubit production. Our results redefine the material limits of superconducting quantum circuits, ensuring that all devices are structurally identical and that control over device performance variability is maintained across the entire wafer and production batches.

## Methods

**Growth of $Al_2O_3$/Al/GaAs(111)A**

The samples were prepared in an ultra-high vacuum (UHV) multi-chamber growth/analysis system, which includes a solid-source GaAs-based III-V compound semiconductor MBE chamber, oxide deposition chambers, a Si/Ge MBE chamber, atomic layer deposition (ALD) reactors, a scanning tunneling microscopy (STM) chamber, and an X-ray photoelectron spectrometry (XPS) chamber. These chambers and reactors are interconnected via UHV modules, maintaining a base pressure of 2 x $10^{-10}$ Torr to preserve pristine sample surfaces during transfers between chambers and reactors.

Epi-ready GaAs(111)A wafers, two inches in diameter, underwent native oxide desorption at 630-650°C in the solid-source III-V MBE chamber under an arsenic overpressure. Subsequently, epitaxial GaAs layers were grown at 550°C with subsequent annealing to 650°C to achieve chemically clean, atomically ordered, and morphologically smooth GaAs(111)A surfaces. Aluminum (Al) films were then deposited onto the *epi*-GaAs surfaces using an effusion cell at a growth rate of 0.33 Å/s. The substrate temperature during deposition was nominally maintained below 0°C, outside the measurable range of the thermocouple. The epitaxial growth of GaAs and Al films was monitored in real time using *in-situ* RHEED with an electron beam energy of 18 keV. Immediately after Al film growth, the samples were transferred under UHV to a dedicated oxide chamber, where 3 nm thick aluminum oxide ($Al_2O_3$) layers were deposited *in-situ* on the freshly grown Al films using e-beam evaporation from a stoichiometric $Al_2O_3$ target. This step was performed to prevent surface contamination and oxidation in forming native oxides ($AlO_x$), critical for preserving the as-grown properties of the aluminum films when samples were removed from the multi-chamber growth/analysis system for subsequent out-of-UHV experiments, analysis, and processing.

**Structural characterization of the Al thin films**

Synchrotron X-ray diffraction (S-XRD) measurements were performed to study the crystallinity and structure of the Al thin films. Measurements were performed at BL09A beamlines of Taiwan Photon Source (TPS), at the National Synchrotron Radiation Research Center (NSRRC), Taiwan, using 9.5 keV incident X-ray at room temperature. Electron backscatter diffraction (EBSD) measurement was carried out with a field emission scanning electron microscope (JSM-7800F Prime) with an EBSD detector (Oxford Instrument NordlysMax[3]), and the electron voltage was controlled at 20 kV. Scanning transmission electron microscopy (STEM) measurement was carried out using an aberration-corrected JEOL 2100F microscope, operating at an accelerating voltage of 200 kV with a 0.9 Å probe size. Specimens for cross-section STEM measurements were obtained using focused ion-beam (FIB, FEI Helios 1200+) milling and *ex-situ* lift-off. All the samples obtained were transferred to copper grids for STEM characterization.

**Hall bar fabrication**

Hall bar regions were patterned using photolithography and etched via inductively coupled plasma-reactive ion etching (ICP-RIE). AZ5214E photoresist was spun at 5800 rpm for 100 seconds for an approximate resist thickness of 1.3 μm and soft baked at 100°C for 2 minutes. Patterning was performed using a mask aligner equipped with near-ultraviolet (NUV) light exposure and TMAH-based developer, followed by rinsing in deionized (DI) water. The dry etching process employed a $BCl_3/Cl_2$ (20sccm/40sccm) mixture at $5.25 \times 10^{-3}$ Torr for $Al_2O_3$/Al. In the same etching sequence, $CF_4$ was subsequently applied to remove $AlCl_x$, followed by a deionized water rinse to dilute any residual $AlCl_x$ after being taken out of the etching system. Photoresist was removed in

1-methyl-2-pyrrolidone (NMP) at 80°C for 30 min with ultrasonic agitation. Contact regions were patterned using photolithography and a lift-off process. Ti/Au (30 nm/100 nm) were deposited by e-beam evaporation and thermal evaporation, respectively. The measured hall bars exhibited a length and width of 400 μm and 100 μm, respectively.

**Preparation of samples for Van der Pauw measurements**

The $Al_2O_3$/Al/GaAs(111)A samples were diced into a square shape with a size of 0.5 cm × 0.5 cm using a diamond scribing wheel. Ti/Au (30 nm/100 nm) was deposited on the four corners through a shadow mask by e-beam evaporation and thermal evaporation, respectively, with each corner pad measuring approximately 0.1 cm × 0.1 cm.

**Electrical transport measurements**

The electrical transport properties were studied using the Hall bar measurements and the four-point van der Pauw method to determine resistivity and superconducting critical temperature, respectively. The temperature dependence of resistivity from 300 K to 2 K was measured using a Physical Property Measurement System (PPMS), and the superconducting transition curves were measured in a He-3 cryostat equipped with electrical measurement instrumentation.

## Data availability

All data are available in the main text or Supplementary Information.
Source data are provided with this paper.

## Acknowledgments

The authors would like to thank the support from the National Science and Technology Council (NSTC) through No. NSTC 114-2112-M-002-027-. The authors acknowledge the STEM technical services of the NTU Consortium of Electron Microscopy Key Technology.

## Author contributions

M. Hong and J. Kwo supervised the project. H.W. Wan and Y.T. Cheng grew the samples; C.K. Cheng characterized the crystalline structure under the supervision of C.H. Hsu; H.W. Wan analyzed the data of EBSD measurements. H.W. Wan and Y.T. Cheng fabricated the Hall-bar samples and performed the electrical transport measurements from 300 K to 2 K with the assistance of R.M. Chia. H.W. Wan and Y.T. Cheng prepared the Van der Pauw samples and performed the

# Supplementary Information

# Near-single-domain superconducting aluminum films on GaAs(111)A with exceptional crystalline quality for scalable quantum circuits

Hsien-Wen Wan[1,#], Yi-Ting Cheng[1,#], Chao-Kai Cheng[1,2], Jui-Min Chia[3], Chien-Ting Wu[4], Sheng-Shiuan Yeh[5], Chia-Hung Hsu[2], Jueinai Kwo[3,*], Minghwei Hong[1,*]

[1] Grad. Institute of Applied Physics and Dept. of Physics, National Taiwan Univ., Taipei, Taiwan

[2] National Synchrotron Radiation Research Center, Hsinchu, Taiwan

[3] Department of Physics, National Tsing Hua University, Hsinchu, Taiwan

[4] Taiwan Semiconductor Research Institute, Hsinchu, Taiwan

[5] Intl. College of Semiconductor Technology, Natl. Yang Ming Chiao Tung Univ., Hsinchu, Taiwan

[#] These authors contributed equally to this work.

* Authors to whom the correspondence should be addressed. Electronic addresses:

raynien@phys.nthu.edu.tw (J. Kwo), and mhong@phys.ntu.edu.tw (M. Hong).

**Supplementary Note 1: The RHEED images of Al thin films**

*In situ* RHEED was employed to monitor the epitaxial growth of Al films on GaAs in real-time. The growth of a GaAs buffer layer on an epi-ready GaAs(111)A substrate resulted in bright, distinct, steaky 2 × 2 RHEED patterns with bright spots and Kikuchi arcs (Fig. S1a), indicating attainment of an atomically ordered and flat surface, ideal for subsequent growth of epi-Al films. Figs. S1 b and c show the RHEED patterns of 9.6 and 19.4 nm thick Al films, respectively, revealing the epitaxial growth of smooth and well-ordered Al films. By monitoring the RHEED patterns of GaAs wafers and Al films throughout the growth process, we determined that the in-plane Al$\langle 1\bar{1}0 \rangle$ aligns with GaAs$\langle 1\bar{1}0 \rangle$. These findings are consistent with the results from SR-XRD measurements and cross-sectional HAADF-STEM on the heterointerface of Al/GaAs(111)A. Moreover, using *in situ* RHEED for studying the growth of Al/GaAs heterostructures allowed us to directly observe atomic order and flatness of the surface in each constituent, the related interface, and the in-plane structural orientation relationship.

The RHEED patterns of the 20 nm thick Al film deposited on sapphire(0001) exhibited relatively diffused and discontinuous streaks (Fig. S1d), indicating limited in-plane crystalline order and surface roughness, as reported by Lin, Y. H. G. *et al.* in *Journal of Applied Physics* **136**, 074401 (2024). In contrast, the RHEED streaks observed for the growth of both 9.6 nm and 19.4 nm thick Al films on GaAs(111)A (Figs. S1 b and c) were significantly sharper and more distinct, reflecting superior crystallinity and smoother surface morphologies. These results highlight the advantages of GaAs(111)A as a substrate for achieving high-quality epitaxial Al films, which are essential for applications requiring precise control over structural and surface properties. The difference in RHEED streak spacing between GaAs and Al along both in-plane directions, $\langle 1\bar{1}0 \rangle$ and $\langle \bar{2}11 \rangle$, corresponds to a lattice mismatch of 28.4%, indicating that the epitaxial Al films were fully relaxed rather than strain-grown on GaAs(111)A.

Nevertheless, excellent epitaxy was observed at the very initial growth stage of Al, as evidenced by the formation of streaky RHEED patterns even at a nominal Al thickness of 0.5 nm (data not shown). Notably, the RHEED streak spacing remained unchanged for Al films with thicknesses of 9.6 nm and 19.4 nm, indicating that the in-plane lattice constants were preserved across this thickness range - consistent with the SR-XRD results.

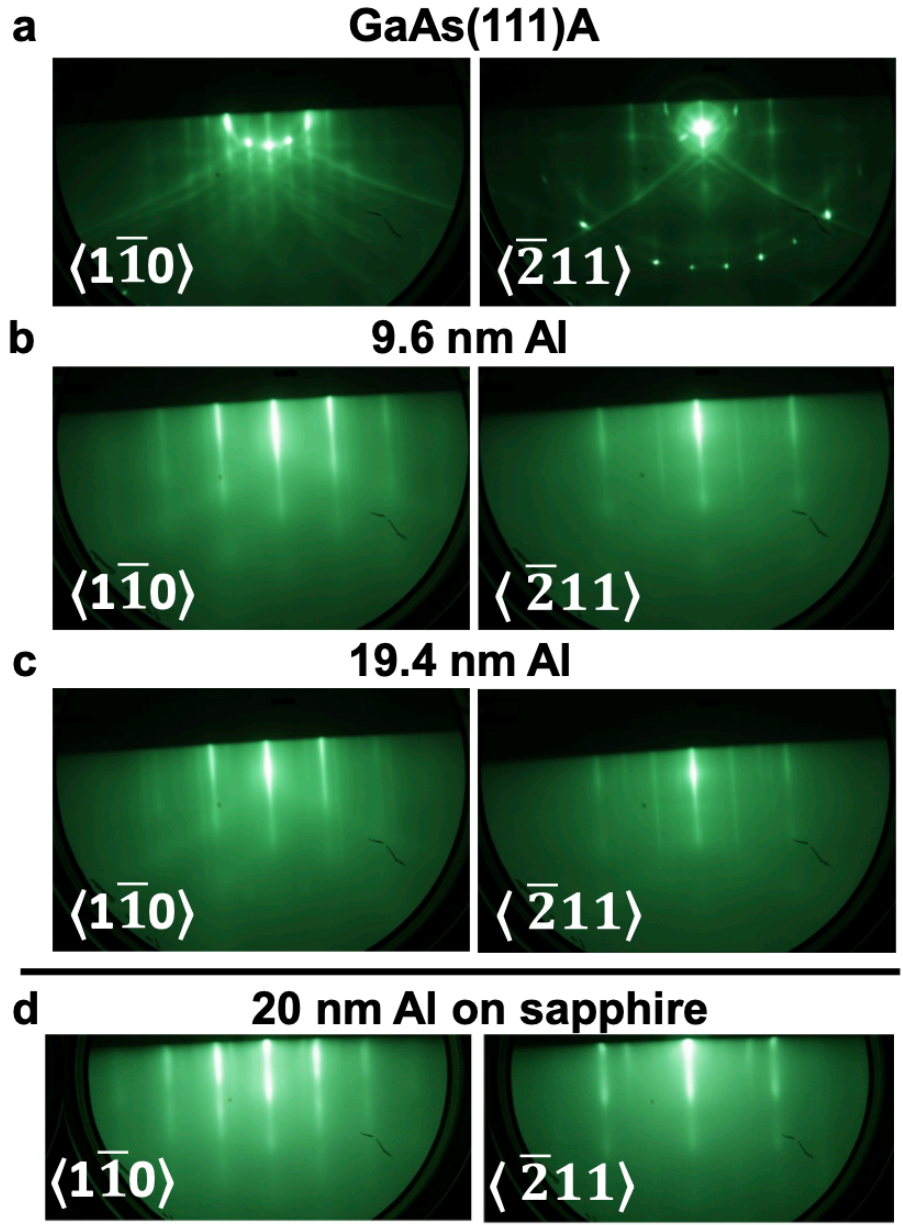

**Fig S1|** *In situ* reflection high-energy electron diffraction (RHEED) images **a**, RHEED of the epitaxial GaAs layer on GaAs(111)A substrate. **b**, 9.6 nm nm thick Al films. **c**, 19.4 nm thick Al films, and **d**, 20 nm thick Al on sapphire from our previous publication, as a comparison. The surface reconstruction of the GaAs(111)A is 2×2 and the in-plane azimuths are $\langle 1\bar{1}0 \rangle$ and $\langle \bar{2}11 \rangle$. The Al film thicknesses were determined by SR-XRD. (Reproduced from Lin, Y. H. G. *et al.* Nanometer-thick molecular beam epitaxy Al films capped with in situ deposited $Al_2O_3$—High-crystallinity, morphology, and superconductivity. *Journal of Applied Physics* **136**, 074401 (2024).)

**Supplementary Note 2: The Atomic Force Microscopy (AFM) Analysis of Film Morphology**

AFM imaging on the *in situ* deposited $Al_2O_3$/Al/GaAs(111)A heterostructures, conducted over a 5 μm × 5 μm scanning area (Figs. S2 a and S2 b), revealed exceptionally flat surface morphologies. The smooth surface morphology, with root mean square roughness ($R_q$) of 0.32 nm, stands in stark contrast to those of Al films deposited on other substrates such as sapphire(0001), Si(111), or GaAs(100) using various techniques including MBE, e-beam evaporation, and sputtering. The notably smooth morphology achieved in our study highlights the importance of selecting appropriate substrates, as well as the necessity of preserving the high quality of these Al thin films throughout the fabrication process. These factors significantly influence the structural perfection of both the metal–substrate and air–oxide–metal interfaces. Such interface quality is essential for minimizing dielectric loss in microwave measurements and for achieving high device performance in Josephson junctions.

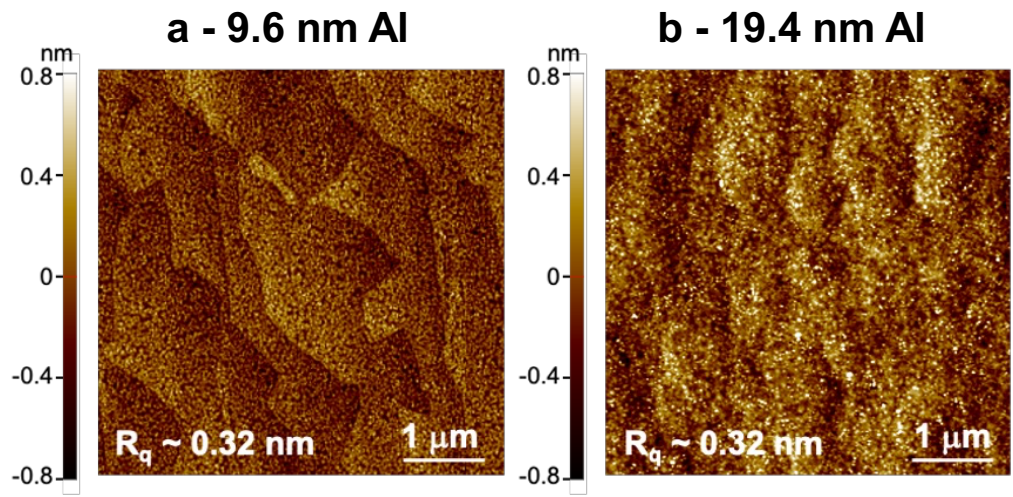

**Figure S2|** Atomic force microscopy images **a,** 9.6 nm thick Al and **b,** 19.4 nm thick Al on GaAs(111)A